\documentclass[aps,showpacs,preprint]{revtex4}
\usepackage[dvips]{color}
\usepackage{amsmath}
\usepackage{epsfig,pstricks,epic,eepic,pst-coil}
\newcommand{\bs}[1]{\ensuremath{\boldsymbol{#1}}}
\begin{document}
\title{Transverse electron scattering response function of $^3$He with $\Delta$-isobar degrees of freedom }

\author{Luping Yuan$^{1}$, Victor D. Efros$^{2}$, Winfried Leidemann$^{1,3}$, 
  and Edward L. Tomusiak$^{4}$}
\affiliation{$^{1}$Dipartimento di Fisica, Universit\`a di Trento,
  I-38100 Trento, Italy\\
$^{2}$ Russian Research Centre "Kurchatov Institute",
  123182 Moscow,  Russia\\
$^{3}$Istituto Nazionale di Fisica Nucleare, Gruppo Collegato
  di Trento, Italy\\
$^{4}$Department of Physics and Astronomy
  University of Victoria, Victoria, BC V8P 1A1, Canada\\
}

\date{\today}

\begin{abstract}
A calculation of the $^3$He transverse $(e,e')$ inclusive response function, $R_T$, which
includes $\Delta$ degrees of freedom is performed using the Lorentz integral transform
method. The resulting coupled equations are treated in impulse approximation, where
the $NNN$ and $NN\Delta$ channels are solved separately.  As NN and NNN potentials
we use the Argonne V18 and UrbanaIX models respectively. Electromagnetic currents include
the $\Delta$-isobar currents, one-body N-currents with relativistic corrections and two-body
currents consistent with the Argonne V18 potential. $R_T$ is calculated for the breakup
threshold region at momentum transfers near 900 MeV/c. Our results are similar
to those of Deltuva {\it et al.} in that large $\Delta$-isobar current
contributions are found. However we find that these are largely canceled by the relativistic
contribution from the one-body N-currents. Finally a comparison is made between
theoretical results and experimental data.

\end{abstract}

\bigskip

%keywords: 
\maketitle

\section{Introduction}

It is well known \cite{RhoWil} that subnuclear degrees of freedom play an important role in 
nuclear dynamics. In conventional low-energy nuclear physics the relevant subnuclear
degrees of freedom are considered to be mesons and nucleon isobars.
Electron scattering affords an excellent tool for studying these degrees of freedom 
which are manifested in the transverse response through meson exchange (MEC)
and isobar (IC) currents.  The consideration of such subnuclear currents has a long history
in the physics of few-nucleon systems. For two-body systems a review can be found
in \cite{ALT05}. One important issue in the MEC is their consistency with the
nucleon-nucleon ($NN$) potential being used. Such consistent MEC's have not only been taken into
account in deuteron electrodisintegration, but also in the electrodisintegration of
three-body systems \cite{ViK00,Golak05,Deltuva1,DEKLOTY08,LEOT10}.
On the other hand IC have not received the attention in three-body systems which they
have in the two-nucleon sector.
Nevertheless there exists a rather complete calculation by \cite{Deltuva1,Yuan02,Deltuva04}
where $N$ and  $\Delta$ degrees of freedom have been treated on an equivalent level via a
coupled channel calculation with $NNN$ and $NN\Delta$ channels. Also $\Delta$-effects
have been studied \cite{ViK00} in $^3$He electrodisintegration below the three-body breakup
threshold using the transition-correlation-operator method \cite{ScW92}.

The present work incorporates the dynamics of the $\Delta$-resonance into the many-body
wavefunction by means of the impulse approximation (IA) \cite{WeA78}.
This method, as in the transition-correlation-operator and coupled channel methods,
avoids the static approximation by fully including the kinetic energy
in the $\Delta$-propagator. For electromagnetic deuteron breakup it has been
shown that the $\Delta$-effects resulting from an IA calculation are rather similar to
those resulting from a coupled channel calculation if the energy is sufficiently below
the resonance position of the $\Delta$ \cite{LA87}. 

A calculation of $R_T$ in our case requires an integration over continuum states of
the coupled NNN + NN$\Delta$ system.  As has been demonstrated previously \cite{ELO94,ELOB07}
the Lorentz integral transform (LIT) method is well suited for calculating inclusive
quantities such as response functions. Examples of its use in calculating electron 
scattering response functions of three- and four-body nuclei employing realistic nuclear forces
(two- and three-body forces) include (i) nonrelativistic calculations of $R_L$
for three-nucleon systems \cite{ELOT04}, for $^4$He \cite{BBLO09} and inclusion of
relativistic effects in $^3$H and $^3$He \cite{ELOT05} and (ii) calculations of $R_T$
with relativistic and consistent MEC contributions for three-body nuclei
\cite{DEKLOTY08,LEOT10,ELOT10}. There is an LIT calculation \cite{BABLO07} of $R_T$ for $^4$He
but with semi-realistic $NN$ forces. The method has not previously been applied to
the coupled NNN+NN$\Delta$ system so that this paper is the first in that regard.

The paper is organized as follows. In section II we describe the general formalism 
including the incorporation of $\Delta$ degrees of freedom in the LIT formalism.
Section III specifies the input to the dynamical equations developed in section II.
This includes subsection A detailing the potentials used in the NNN and NN$\Delta$
sectors, subsection B describing the electromagnetic current operators used and
subsection C outlining calculational details. Finally our results are discussed in
section IV. There we compare our results for $R_T$ to those from another calculation
and to experimental data.

\section{Formalism}

In the one photon exchange approximation
the cross section for the process of inclusive electron scattering on a 
nucleus
is given by
\begin{eqnarray}
{d^2\sigma\over d\Omega\,d\omega}\ =\ \sigma_{Mott}\ \bigg[\ {Q^4\over 
q^4}\,
R_L(q,\omega)\ +
\ \left(\ {Q^2\over 2q^2}+\tan^2{\theta\over 2}\ 
\right)\,R_T(q,\omega)\bigg] \,,
\label{sigma}\end{eqnarray}
where $R_L$ and $R_T$ are the longitudinal and transverse response functions
respectively, $\omega$ is the electron energy loss, $q$ is the magnitude 
of the
electron momentum transfer, $\theta$ is the electron scattering angle, and
$Q^2 = q^2 - \omega^2$.

In the present work we study the transverse response function,
\begin{equation}
R_T(q,\omega)={\overline \sum}_{M_i}\sum\!\!\!\!\!\!\!\int\,df
({\bf j}_t^\dag)_{if}\cdot({\bf j}_t)_{fi}\delta(E_{\bar f}-E_i-\omega) 
\label{resp}
\end{equation}
with $\Delta$ degrees of freedom within a non-relativistic approach.  The
low $\omega$ region is considered. In (\ref{resp}) the subscripts $i$ 
and $f$ label, respectively, an initial state  and
final states, and the matrix elements are taken between internal states, 
center-of-mass motion being excluded.

As mentioned in the introduction we employ the IA in order to take into account the
$\Delta$-resonance. This approximation is used for both the $^3$He ground state and
the final state.  Below we outline the various theoretical aspects required to include
$\Delta$ degrees of freedom in a calculation of $R_T$ via the LIT method.

%\subsection{Hamilton operator and dynamic equations}
We consider the three-nucleon system with $N(939)$ and $\Delta(1232)$ degrees of freedom,
which leads to the following Hamiltonian
\begin{equation}
%H = T + \delta m + V = \sum _{i=1}^3 (T_i + \delta m _i) + \sum ^3 _{i < j} \upsilon_{ij}
H = T + \delta m + V = \sum _{i=1}^3 (T_i + \delta m _i) + \sum ^3 _{i < j} V_{ij} \,,
\end{equation}
where $T_i $ is the kinetic energy of particle $i$ with mass $m_i$, $\delta m _i = m_i - m_N$
is its mass difference with nucleon $N(939)$, and $ V_{ij} = V_{ji}$ is the potential between
particles $i$ and $j$. By omitting the contribution from more than one $\Delta $-isobar
excitation, we construct the three-particle bound state from an $NNN$ part and a $NN\Delta$ part i.e.,
\begin{equation}
|\Psi  _0 \rangle  =    |\Psi ^N _0\rangle +    |\Psi ^{\Delta} _0 \rangle \,.
\end{equation} 
The wave function is determined by the Schr\"odinger equation
\begin{eqnarray}
( T_N + \bar V^{NN} - E) |\Psi ^N _0 \rangle &=& - V^{NN,N\Delta} \ |\Psi ^{\Delta} _0 \rangle \,, \label{schrn} \\    
(H_{\Delta} - E) |\Psi ^{\Delta} _0 \rangle =
( \delta m + T_{\Delta} + V^{N\Delta} - E) |\Psi ^{\Delta} _0 \rangle &=& - V^{N\Delta,NN}  |\Psi ^N _0 \rangle \,, \label{schrd}     
\end{eqnarray}    
where $V^{N'_1 N'_2,N_1 N_2} = \sum ^3 _{i < j} V_{ij} (N_1 N_2 \rightarrow {N'_1 N'_2}),
\quad \bar V^{NN} = \bar V^{NN,NN}, \quad V^{N\Delta} = V^{N\Delta,N\Delta} $, and
$\sum^3_{i=1} T_i$ is denoted by $T_N$ and $T_\Delta$ for the $NNN$ and $NN\Delta$ channels,
respectively. One should note that $\bar V^{NN}$ is different from the usual
 sum of realistic $NN$ potentials, $V^{NN}$,  because the latter already contain
implicit  effects due to the $\Delta$
(e.g., in meson theoretical $NN$ potentials realized via part of the $\sigma$ meson exchange). 

Here we do not search for a direct solution of the coupled channel problem represented
by Eqs.~(\ref{schrn},\ref{schrd}). Instead we use the IA  where
one computes $\Psi ^N _0$  and $\Psi^{\Delta}_0$ separately. More specifically one first
determines the $NNN$ part by solving 
\begin{equation}
(H_N - E)  |\Psi ^N _0 \rangle = 0  \label{newschrn}
\end{equation}
with
\begin{equation}
 H_N=T_N + V^{NN} +  V^{NNN}  \label{H_N} \,,
\end{equation}
where $V^{NNN}=\sum ^3 _{i<j<k} V_{ijk}$ is a three-nucleon force. In the IA one then uses
the solution $|\Psi^N_0 \rangle$ in order to calculate $|\Psi ^{\Delta} _0 \rangle$ through (\ref{schrd}).

Treatment of the continuum in the LIT technique requires the calculation
of a localized Lorentz state $|\tilde\Psi\rangle$.  
This Lorentz state $|\tilde\Psi\rangle$ also has  $NNN$ and $NN\Delta$ parts written as 
\begin{equation}
\label{lorentzstate}
|\tilde\Psi  \rangle =    |\tilde\Psi ^N  \rangle +    |\tilde\Psi ^{\Delta} \rangle \,.
\end{equation}
These fulfill the coupled equations
\begin{eqnarray}    
 ( T_N + \bar V^{NN} - E_0 - \sigma ) |\tilde\Psi ^N  \rangle &=& - V^{NN,N\Delta} \ |\tilde\Psi ^{\Delta}
 \rangle + {O}_{NN}  |\Psi ^N _0 \rangle  + {O}_{N\Delta}  |\Psi ^{\Delta} _0 \rangle \,, \label{schrlorn}\\
\label{lorentzstate2} ( \delta m + T_{\Delta} + V^{N\Delta} - E_0 -\sigma) |\tilde\Psi ^{\Delta} 
 \rangle &=& - V^{N\Delta,NN}  |\tilde\Psi ^N  \rangle  + {O}_{\Delta N}  |\Psi ^N _0 \rangle
  + {O}_{\Delta \Delta}  |\Psi ^{\Delta} _0 \rangle \label{schrlord} \,,
\end{eqnarray}    
where $E_0$ is the three-body ground-state energy, the complex $\sigma=\sigma_R + i \sigma_I$
 is the argument of the LIT in the transformed space, and the ${O}_{N_1 N_2}$ denote the
 various diagonal ($N_1=N_2)$) and transition ($N_1 \ne N_2)$ electromagnetic current operators.
One first solves for the $NNN$ part using  $H_N$ of Eq.~(\ref{H_N}):
\begin{eqnarray}
\label{lorentzstate1} ( H_N  - E_0  - \sigma ) |\tilde\Psi ^N  \rangle &=& - V^{NN,N\Delta}
 ( H_{\Delta}  - E_0 - \sigma)^{-1} \left( {O}_{\Delta N}  |\Psi ^N _0 \rangle  + {O}_{\Delta\Delta}
  |\Psi ^{\Delta} _0 \rangle \right)  \nonumber \\
&& + {O}_{NN}  |\Psi ^N _0 \rangle  + {O}_{N\Delta}  |\Psi ^{\Delta} _0 \rangle \label{newschrlorn} \,.
\end{eqnarray}
The above equation is derived by solving (\ref{lorentzstate2}) formally for $|\tilde\Psi^{\Delta}\rangle$,
inserting the solution in (\ref{schrlorn}), and dropping the term $V^{NN,N\Delta} ( H_{\Delta} - E_0 - \sigma)^{-1}
V^{N\Delta,NN}$, since, as mentioned, $\Delta$-effects to the nuclear interaction are already contained in the realistic $H_N$.
With $ |\tilde\Psi ^N  \rangle $ thus obtained one then calculates $|\tilde\Psi ^{\Delta}  \rangle$
 in a second step through (\ref{schrlord}).
Given the solutions $|\tilde\Psi^N\rangle$ and $|\tilde\Psi^\Delta\rangle$ the LIT is obtained from
the norm of the Lorentz state as
\begin{equation}
 \langle\tilde\Psi|\tilde\Psi\rangle\  =\ \langle\tilde\Psi^N|\tilde\Psi^N\rangle +
 \langle\tilde\Psi^\Delta|\tilde\Psi^\Delta\rangle.
\end{equation}
 These two terms correspond to different
 contributions. It can be shown that the piece $\langle\tilde\Psi^N|\tilde\Psi^N\rangle$
 describes contributions due to final states with nucleons only. 
In this case the $\Delta$ degrees of freedom 
 only contribute as virtual intermediate states. On the contrary the term 
$\langle\tilde\Psi^\Delta|\tilde\Psi^\Delta\rangle$ describes contributions from final states
containing a real $\Delta$.  The contribution to $R_T$ from this term vanishes below the threshold
for $\Delta$ production.  Since the present study is for energies below 
that threshold this term will not contribute here to $R_T$.
Such a real $\Delta$ has to decay into a nucleon and a pion eventually,
 thus the contribution $\langle\tilde\Psi^\Delta|\tilde\Psi^\Delta\rangle$ corresponds to resonant
 pion production.

\section{Input to Dynamical Equations}
\subsection{Potentials}

In the pure nucleonic sector we use the Argonne V18 (AV18) NN potential \cite{AV18} and the
Urbana IX (UIX) NNN potential \cite{UIX} while for the pure NN$\Delta$ sector we
take $V^{N\Delta}$=0 as in the IA calculation of \cite{LA87}. The NNN and
NN$\Delta$ sectors are coupled via the $V^{NN,N\Delta}$ and $V^{N\Delta,NN}$ potentials.
We use the same form for this potential as described in \cite{LA87} except that here we use
the short range cutoff given in the AV18 \cite{AV18} potential.
In detail we take $V(NN\to N\Delta)$ between particles 1 and 2 to have the form
\begin{eqnarray}
V_{12} = a_{NN'\pi}{\bs \tau}_1 \! \cdot \!  {\bs \tau}_2 \left \{ \left [ V _0(m_{\pi}) + \frac{2a_{NN'\rho}}{a_{NN'\pi}}V _0(m_{\rho}) \right ] {\bs \sigma}_1 \! \cdot \!  {\bs \sigma}_2 \right . \nonumber \\
\left . + \left  [ V _2(m_{\pi}) - \frac{ a_{NN'\rho}}{a_{NN'\pi}} V _2(m_{\rho}) \right ] S_{12} \right \}
\end{eqnarray}
with
\begin{equation}
V _0(m) = (m r)^{-1} e^{-m r}(1-e^{-cr^2}),
\end{equation}
\begin{equation}
V _2(m) = \left [ 1+ \frac{3}{mr} + \frac{3}{(mr)^2 }\right ]\frac{e^{-mr}}{mr}(1-e^{-cr^2})^2,
\end{equation}
where $ {\bs \sigma}_i$ $({\bs \tau}_i)$ are regarded as transition operators for
spin (isospin) of particle $i$, the coupling constants $a_{NN'\pi}$
and $a_{NN'\rho}$ are taken from
\cite{LA87}, $c = 2.1 \; \mbox{fm} ^{-2}$ is the same value as in the AV18 potential,
and $r = |{\bf r}_1 - {\bf r}_2|$ is proportional to the Jacobi vector ${\bs{\xi}}_1$
(see Eq.~(\ref{xidef})).
% =  [(A_1+A_2)/A_1A_2]^{1/2}\xi _1 $.

\subsection{Electromagnetic Currents}

In Eqs.~(\ref{schrlorn}) and (\ref{schrlord}) the driving terms are the transverse 
electromagnetic currents acting on the ground state.  The term $O_{NN}$ represents
the purely nucleonic one- and two-body currents.  For these the same one-body and
two-body currents as employed in \cite{ELOT10} are used.
There the one-body currents included relativistic corrections to order $M^{-3}$ and the
two-body currents were consistent $\pi$- and $\rho$-MEC currents constructed using
the method of Arenh\"ovel and Schwamb \cite{AS}. The other terms $O_{N\Delta}$,
$O_{\Delta N}$, and $O_{\Delta\Delta}$ are transition and diagonal
one-body $\Delta$-isobar currents. The currents involving the $\Delta$-resonance are given in Fig.~1.  For one-body $\Delta$-isobar currents we use the forms
\begin{equation}
{\bf j}_{\Delta N}=\ \sum_{k=1}^3
e^{i{\bf q}\cdot{\bf r}_k^\prime}\ \frac{i({\sigma_k}^{\Delta N}\times{\bf q})}{2m_{\Delta}}\ G^{\Delta N}_{M1}\tau_{zk}^{\Delta N},
\end{equation}
and
\begin{eqnarray}
{\bf j}_{\Delta \Delta}=\ \sum_{k=1}^3
e^{i{\bf q}\cdot{\bf r}_k^\prime}\ \left [\frac{2{\bf p}_k^\prime + \kappa{\bf q}}{2m_{\Delta}}\  G^{\Delta}_{E0} +\ \frac{i({\sigma_k}^{\Delta }\times{\bf q})}{2m_{\Delta}}\  \ G^{\Delta }_{M1}\right ]\frac{1+\tau_{zk}^{\Delta}}{2},\label{dd}
\end{eqnarray}
where ${\bf r}_k^\prime = {\bf r}_k-{\bf R}_{cm}$,
 and ${\bf p}_k^\prime = {\bf p}_k-{\bf P}_{cm}A_k/A$ are the relative
coordinate and momentum operator of the k-th particle, while  ${\bf R}_{cm}$ and ${\bf P}_{cm}$ are the center-of-mass
coordinate and initial total momentum variables of the system. 
With the assumption that ${\bf P}_{cm}$ is directed along $\bf q$ the term $\kappa$  
 in (\ref{dd}) has the value $\kappa = 1 + 2P_{cm}A_k/(Aq)$. However since here we are
dealing with transverse currents this term does not contribute. 
The formfactors for the above currents are the same as in \cite{Deltuva1,Deltuva04}
and take the form
\begin{eqnarray}
% G^{\Delta N}_{M1}(Q^2) = \frac{6.366}{1+Q^2/{\Lambda}^2_{\Delta N}}, \;  G^{\Delta}_{E0}(Q^2)  = G^p_E(Q_2), \;  \ G^{\Delta }_{M1}(Q^2)=(m_{\Delta}/3m_N)G^p_M,
&& G^{\Delta N}_{M1}(Q^2) = {\frac {m_{\Delta}}{m_N} } \frac{4.59}{(1+Q^2/{\Lambda}^2_{\Delta N1})^2(1+Q^2/{\Lambda}^2_{\Delta N2})^{1/2}}, \nonumber \\
&&  G^{\Delta}_{E0}(Q^2)  = G^p_E(Q_2), \\
&& G^{\Delta }_{M1}(Q^2)= {\frac {m_{\Delta}}{3m_N} } \frac{(4.35/2.0)}{(1+Q^2/{\Lambda}^2_{\Delta N1})^2} \nonumber
\end{eqnarray}
%with ${\Lambda}_{\Delta N}= 841.8$ MeV.
with ${\Lambda}_{\Delta N1}=  840 $ MeV and ${\Lambda}_{\Delta N2}= 1200$ MeV .
As in our previous $NNN$ calculations \cite{DEKLOTY08,LEOT10,ELOT10} we use the approximations
\begin{eqnarray}
G_{M1}^{\Delta N}(Q^2)\approx{\bar\mu}^{\Delta N}_p(Q_{av}^2)G_E^p(Q^2),\qquad\qquad
{\bar\mu}^{\Delta N}_p(Q_{av}^2)=\frac{G^{\Delta N}_{M1}(Q_{av}^2)}{G_E^p(Q_{av}^2)}, \nonumber\\
G_{M1}^{\Delta}(Q^2)\approx{\bar\mu}^{\Delta}_p(Q_{av}^2)G_E^p(Q^2),\qquad\qquad
{\bar\mu}^{\Delta}_p(Q_{av}^2)=\frac{G_{M1}^{\Delta}(Q_{av}^2)}{G_E^p(Q_{av}^2)},\nonumber
\end{eqnarray} .

\begin{figure}[!]
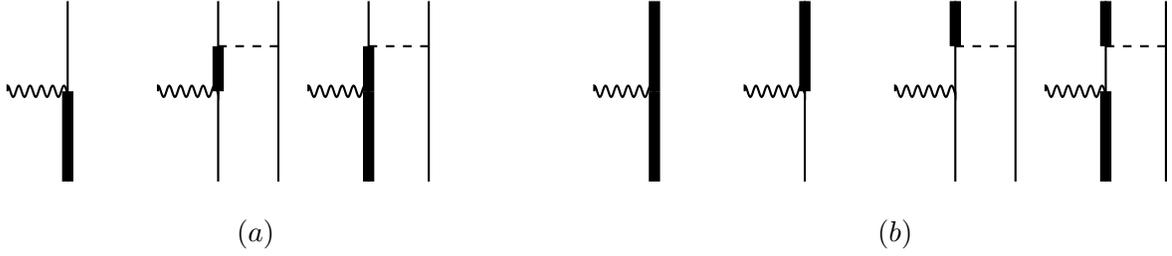

\begin{center}
\pspicture(15.0,3.0)
\def\nucleon{\psline(0,0)(0,2.4)}
\def\NNPINN{\psline(0,0)(0,2.4)\psline(0.8,0)(0.8,2.4)
\psline[linestyle=dashed,dash=3pt 3pt](0,1.8)(0.8,1.8)}
\def\NNNN{\psline(0,0)(0,2.4)\psline(0.8,0)(0.8,2.4)}
\def\photon{\pscoil[coilwidth=0.15cm,coilaspect=0,coilarm=0.0cm](0,0)(0.8,0)}
\def\deltaiso{\psline[linewidth=0.15cm](0,0)(0,1.2)}
\def\shortdeltaiso{\psline[linewidth=0.15cm](0,0)(0,0.6)}
\rput(1,0){\nucleon}
\multips(3,0)(2,0){2}{\NNPINN}
\multips(1.0,0.0)(4,0){2}{\deltaiso}
\multips(3,1.2)(2.0,0){2}{\shortdeltaiso}
\multips(0.2,1.2)(2,0){3}{\photon}
\rput(3.5,-0.7){$(a)$}
% next 4
\multips(8.0,1.2)(2,0){4}{\photon}
%\multips(9.8,0)(2,0){1}{\NNNN}
\rput(10.8,0){\nucleon}
\multips(12.8,0)(2,0){2}{\NNPINN}
\multips(8.8,0.0)(0,1.2){2}{\deltaiso}
\multips(10.8,1.2)(0,1.2){1}{\deltaiso}
\multips(12.8,1.8)(2,0){2}{\shortdeltaiso}
\multips(14.8,0)(2,0){1}{\deltaiso}
\rput(12,-0.7){$(b)$}
\endpspicture
\end{center}
\caption{\label{fig:Jnd}
Operators involving the $\Delta$-resonance in the $NNN$ (a) and $NN\Delta$ (b) channels}
\end{figure}

\subsection{Calculational Details}

In order to solve Eqs.~(\ref{newschrn}), (\ref{schrlord}) and (\ref{newschrlorn}) for the
ground state and Lorentz vectors we expand the bound and Lorentz
states on a complete antisymmetric basis. The reason for antisymmetrizing  NN$\Delta$ states is that 
they couple to purely antisymmetric nucleonic states
through symmetric operators.  Thus the excitation of an antisymmetric NNN state
to a $NN\Delta$ state occurs via an operator symmetric with respect to nucleons.
Such an operator is a sum of operators which replace
a nucleon with a $\Delta$ which therefore
leads to an antisymmetric $NN\Delta$ state. For the $NNN$ part we take the same correlated
hyperspherical basis as in our previous three-body calculations without
considering $\Delta$ degrees of freedom (see, e.g., \cite{ELOB07}). For
the part with one $\Delta$-excitation we use the following hyperspherical basis
\begin{equation}\label{basis1}|\varphi _k \rangle = \frac{1+P}{\sqrt{3}}\frac{1-(12)}{\sqrt{2(1+B^2)}}
\left|\left(h_{NK(Ll)\mathcal{L}} \otimes \left(\left(s_1s_2\right)Ss_3\right)
\mathcal{S}\right)_{\mathcal{JM_J}}, \left(\left(t_1t_2\right)Tt_3\right)\mathcal
{TM_T}\right \rangle \,,
\end{equation}
where $N$ is the order of the hyperradial function, $K$ is the hyperspherical
angular momentum, $L$ and $l$ are the orbital angular momentum of the pair and
of the spectator, respectively, coupled to the total orbital angular momentum
$\mathcal{L}$. Individual spin (isospin) quantum numbers 
of the three particles are denoted by $s_i$ $(t_i),$ $i = 1,2,3$ while the pair spin
(isospin) is denoted by $S$ ($T$), the total spin (isospin) by $\mathcal{S}$
($\mathcal{T}$) and $\mathcal{M_T}$ stands for the projection of the total
isospin. Quantum numbers ${\mathcal J}$ and ${\mathcal M_J}$ denote the total
angular momentum and its projection and $(...\otimes...)_{\mathcal JM_J}$ denotes
$\mathcal{(LS)J}$ coupling. The index $k$ denotes collectively
$ \{N,K,L,l,\mathcal{L},S,\mathcal{S},{\mathcal {J,M_J}},T,\mathcal{T,M_T} \}$
and $B$. We define the quantity $B$ to be 0 if the pair of the three particle
system contains one $\Delta$ and therefore $(s_1,s_2,s_3) = (t_1,t_2,t_3)=(1/2,3/2,1/2)$,
and to be 1 if the pair contains no $\Delta$ and therefore
$(s_1,s_2,s_3) = (t_1,t_2,t_3)=(1/2,1/2,3/2)$. Note that if particles 1 and 2
are nucleons we always assume that L+S+T=odd so that the NN pair is already
antisymmetric.
The spatial basis functions in coordinate representation are
products of hyperradial functions and hyperspherical harmonics 
\begin{equation}
\langle {\bs \xi}_1{\bs \xi}_2 | h_{NK(Ll)\mathcal{LM_L}}\rangle = R_N(\rho)Y^{Ll}_{K\mathcal{LM_L}}
(x,x_1,\varphi _1,x_2,\varphi _2),
\end{equation}
The coordinates $\rho$ and $x$  are defined
in terms of the Jacobi vectors
\begin{eqnarray}
{\bs{\xi}}_1 &=& \sqrt{\frac{A_1A_2}{A_1+A_2}}\left({\bf r}_2 - {\bf r}_1\right),
\;\;{\bs{\xi}}_2 = \sqrt{\frac{(A_1+A_2)A_3}{A_1+A_2+A_3}}\left({\bf r}_3 - \frac{A_1
{\bf r}_1 + A_2 {\bf r}_2}{A_1+A_2}\right) \label{xidef}
\end{eqnarray}
as $\rho=(\xi_1^2+\xi_2^2)^{1/2}$, $x=(\xi_2^2-\xi_1^2)/\rho$.  The  coordinates
$x_i \equiv \cos \theta _i$ and $\varphi _i$  are
spherical coordinates of the unit vectors in the directions of ${\bs{\xi}}_1$ and ${\bs{\xi}}_2$. 
 We use the notation $A_i \equiv {m_i}/{m_N} $. Note that particle permutations
entering the antisymmetrization operator
interchange not only particle position vectors ${\bf r}_i$ but also their mass numbers $A_i$. One has 
\mbox{$\rho^2=\xi_1^2+\xi_2^2=A_1r_1^2+A_2r_2^2+A_3r_3^2-AR^2$}, where
${\bf R}$ is the center-of-mass position. Thus $\rho$ remains invariant under 
particle permutations.

The operator $\left(1-(12)\right)/\left[2(1+B^2)\right]^{1/2}$ makes the (12)-pair explicitly
antisymmetric. Note that it gives unity if the pair contains no $\Delta$ but rather is an
antisymmetric NN pair. 
Finally the operator ${(1+P)}/{\sqrt{3}}$, where $P \equiv (123) + (132)$, makes the three
particle states with antisymmetric (12) pair totally antisymmetric.
It turns out to be convenient to keep both $B = 0$ and $B = 1$ in
(\ref{basis1}) (thereby resulting in an overcomplete basis) and to
finally select out numerically the linearly independent states.
This enables one to select out those states which give negligible contributions to the results. 
Application of the operator ${(1+P)}/{\sqrt{3}}$ in (\ref{basis1}) 
results in the more practical form 
\begin{equation}
|\varphi _k \rangle = \sum _{B'S'T'}\frac{1-(12)}{\sqrt{2(1+B^{'2})}}\left|\left(F_{j \mathcal{
L}}\otimes \left(\left(s_1's_2'\right)S's_3'\right)\mathcal{S}\right)_{\mathcal{JM_J}}
,\left(\left( t_1't_2'\right)T't_3'\right)\mathcal{TM_T}\right \rangle, \label{phik}
\end{equation}
where
$j$ denotes collectively $\{ B'S'T',BS\mathcal{S}T\mathcal{T}NK(Ll)\}$ and
\begin{eqnarray}\left | F_{j \mathcal{L}\mathcal{M_L}}\right \rangle &&= \frac{1}{\sqrt{3}} \Big ( \delta _{B',B}  \delta _{S',S}  \delta _{T',T}  \left | h_{NK(Ll)\mathcal{L}\mathcal{M_L}}\right \rangle 
\nonumber \\ && + g ^c _{B'S'T',BS\mathcal{S}T\mathcal{T}}\left | h^c_{NK(Ll)\mathcal{L}\mathcal
{M_L}} \right \rangle + g ^d _{B'S'T',BS\mathcal{S}T\mathcal{T}} \left | h^d_{NK(Ll)\mathcal{L}\mathcal{M_L}} \right \rangle \Big ) \label{F}\end{eqnarray}
with $(Ll)\mathcal{L} $ coupling for the total orbital angular momentum $\mathcal{L}$ and
 its projection $\mathcal{M_L}$.
As mentioned above components with $B'$=1 in Eq.~(\ref{phik}) represent configurations
NN$\Delta$ in which particles 1 and 2 are nucleons. Those components with $B'$=0 represent
N$\Delta$N configurations in which particle 1 is a nucleon and particle 2 is a $\Delta$.
 More details of the spin-isospin factors $g^c$ and $g^d$, and the
 spatial functions $h^c$ and $h^d$ are given in Appendix \ref{gh}.
Techniques employed in calculating the kinetic energy and the
$NN\to N\Delta$ or $N\Delta\to NN$ potential are given in Appendices \ref{KE} and
\ref{pot} respectively. 

As in \cite{DEKLOTY08} all currents are expressed in terms of multipole 
expansions. Explicit expressions for the multipoles of the one-body current
(containing relativistic corrections) are given in \cite{ELOT10}. 
The multipoles for the $\pi$- and $\rho$-MEC are found in \cite{DEKLOTY08}
with modifications due to the implementation of consistent MECs for the
AV18 potential listed in \cite{LEOT10}. Finally the multipoles required
here for the one-body currents relating to the $\Delta$ are listed in
Appendix \ref{multidelta}. With these multipoles one can then decompose the
LIT of the response function according to its multipole content as
\begin{equation}
\label{rcomp}
\tilde{R_T}(\sigma)=
\frac{4\pi}{2{\mathcal J}_0+1}\sum_{\lambda={\rm el,mag}}\sum_{{\mathcal J}j}
(2{\mathcal J}+1)(\tilde{R_T})_{\mathcal J}^{j\lambda} \,,
\end{equation}
where 
\begin{equation}
\label{tilder}
(\tilde{R_T})_{\mathcal J}^{j\lambda}=\langle \tilde{\Psi}^{j\lambda}_{\mathcal JM}
 | \tilde{\Psi}^{j\lambda}_{\mathcal JM} \rangle = \langle \tilde{\Psi}^{Nj\lambda}_
{\mathcal JM} | \tilde{\Psi}^{Nj\lambda}_{\mathcal JM} \rangle + \langle
 \tilde{\Psi}^{\Delta j\lambda}_{\mathcal JM} | \tilde{\Psi}^{\Delta j\lambda}_{\mathcal JM}
 \rangle.
\end{equation}
Here ${\mathcal J}_0$ is the initial state total angular momentum, 
${\mathcal J}$ and ${\mathcal M}$ are the final state total
 angular momentum and its projection.  The  $| \tilde{\Psi}^{j\lambda}_{\mathcal J}
 \rangle$ are the solutions of (\ref{lorentzstate2},\ref{lorentzstate1}) 
where the following replacement is made on the rhs of these equations
\begin{equation} \label{OPsi} 
O|\Psi _0 \rangle \rightarrow |q_{{\mathcal J}{\mathcal M}}^{j\lambda}\rangle =
 [T_j^\lambda \otimes |\Psi_0({\mathcal J}_0)\rangle]_{{\mathcal J}
{\mathcal M}}. 
\end{equation}
In Eq.~(\ref{tilder}) ${\mathcal M}$ is arbitrary.
In Eq.~(\ref{OPsi}) above $\Psi _0$ is either $\Psi ^N _0$ or $\Psi^{\Delta} _0$,
while $O$ represents the various electromagnetic current operators on the
rhs of (\ref{schrlorn},\ref{lorentzstate2}).
By projecting the rhs of (\ref{OPsi}) on the basis states (\ref{phik}) one obtains
\begin{equation}
\label{basisq}
  \left \langle \left (F_{j'\mathcal{L'}}\otimes \left(\left(s_1's_2'\right)S's_3'\right)\mathcal{S'}\right)_{\mathcal{J'M'}} \right |q_{{\mathcal J}{\mathcal M}}^{j\lambda}\rangle = 
  \delta_{\mathcal{J',J}}\delta_{\mathcal{M',M}}\frac{ \left \langle \left (F_{j'\mathcal{L'}}\otimes \left(\left(s_1's_2'\right)S's_3'\right)
  \mathcal{S'}\right)_{\mathcal{J'}} \right \|T_j^\lambda \| \Psi_0({\mathcal J}_0)\rangle}{\hat{\mathcal J'}} .
\end{equation}
We use the Lanczos method to calculate the response function, as described in \cite{MBLO03}.  The response function is thus calculated by using
\begin{eqnarray}
\label{imtilder}
 \langle \tilde{\Psi}^{Nj\lambda}_{\mathcal {JM}} | \tilde{\Psi}^{Nj\lambda}_{\mathcal {JM}} \rangle = -\frac{1}{\sigma _I}Im  \langle Q^{j\lambda}_{\mathcal{JM}} |\frac{1}{E_0 + \sigma - H_N}  |  Q^{j\lambda}_{\mathcal{JM}}  \rangle  \nonumber \\
= -\sum _{m,n}\frac{1}{\sigma _I}Im  \langle Q^{j\lambda}_{\mathcal{JM}} | \phi _m \rangle \langle \phi _m |  \frac{1}{E_0 + \sigma - H_N}  | \phi _n \rangle \langle \phi _n |   Q^{j\lambda}_{\mathcal{JM}} \rangle \,,
\end{eqnarray}
where $Q$ corresponds to the rhs of (\ref{lorentzstate1}) and $| \phi _n \rangle $ is the set of orthogonal Lanczos vectors. As starting vector $| \phi _0 \rangle $ we choose the rhs of (\ref{lorentzstate1}) at one particular value of $\sigma $. The expression $\langle \phi _m |  ({E_0 + \sigma - H_N})^{-1}  | \phi _n \rangle $ can be calculated by using Eq.~(71) of \cite{MBLO03}. A difference from previous LIT applications appears in the potential term on the rhs of the LIT equation, namely the $N\Delta$ transition potential $V^{NN,N\Delta}$ in (\ref{lorentzstate1}).
The contribution of this term to $\langle \phi _n |   Q^{j\lambda}_{\mathcal{JM}} \rangle $ 
is given by
\begin{equation}
-\sum_{l'k'lkm}\langle \phi _n  | \varphi _{l'} \rangle N^{-1}_{l'k'}\langle \varphi _{k'} | V^{NN,N\Delta}  | \varphi _{l} \rangle N^{-1}_{lk}\langle \varphi _{k} | \phi _m \rangle \langle \phi _m |( H_{\Delta}  - E_0 - \sigma)^{-1} |q^{j\lambda}_{\mathcal{JM}} \rangle,
\end{equation}
with
\begin{equation}
|q^{j\lambda}_{\mathcal{JM}} \rangle = [{\cal T}_{\Delta N,  j}^\lambda \otimes |\Psi^N_0({\mathcal J}_0)\rangle]_{{\mathcal J}{\mathcal M}}+  [{\cal T}_{\Delta \Delta,  j}^\lambda \otimes |\Psi^{\Delta}_0({\mathcal J}_0)\rangle]_{{\mathcal J}{\mathcal M}}, 
\end{equation}
and the norm matrix $N = \left( \left \langle \varphi _{k} | \varphi _{l} \right \rangle  \right)$. 
Also the second term in Eq.~(\ref{tilder}), $\langle \tilde{\Psi}^{\Delta j\lambda}_{\mathcal JM} | \tilde{\Psi}^{\Delta j\lambda}_{\mathcal JM} \rangle $, can be calculated in a similar way with the Lanczos method.

Because in this paper we are working at low energies near the $^3$He breakup
threshold only the lowest multipole transitions contribute. We found sufficient
accuracy by restricting the maximum value of $\mathcal J$ to 5/2.
The LIT is computed with $\sigma_I=5$ MeV. The LIT inversion \cite{BELO10} is
made with our standard inversion method \cite{ELOB07,ALRS05}.
As discussed in \cite{DEKLOTY08} we subtract from the LIT of the M1
transition the elastic contribution and invert the remaining inelastic piece
separately from the other multipole contributions.

\section{Results}
In the present work we have used the LIT method to calculate $\Delta$-IC effects on
the transverse electron scattering response function $R_T(q,\omega)$ for
q$\approx$900 MeV/c and $\omega$ up to 20 MeV above the breakup threshold. This is the first application of the
LIT method to include $\Delta$ degrees of freedom in the calculation of inclusive
(e,e') response functions.
The importance of $\Delta$-effects at these kinematics has previously
been shown by \cite{Deltuva1}.
As $NN$ and $NNN$ interactions we used the AV18 and the UIX potentials respectively. Following the IA calculation of
\cite{LA87} we do not consider a diagonal N$\Delta$ interaction, {i.e.} $V_{N\Delta}=0$.
For the $^3$He ground state, our interaction model leads to a $\Delta$-probability of 1.14 \%
which compares to 1.44\% obtained by \cite{Deltuva1} who used a CDBonn+$\Delta$ coupled
channel potential model \cite{DeM03}.  In addition to the $\Delta$-ICs ${\bf j}_{N\Delta}$,
${\bf j}_{\Delta N}$ and ${\bf j}_{\Delta \Delta}$ the purely nucleonic currents
include the nonrelativistic one-body current with relativistic corrections up to 
order $M^{-3}$ \cite{ELOT10} and an MEC consistent with the AV18 potential \cite{LEOT10}.
Concerning the relativistic corrections we leave out the $\omega$ dependent relativistic piece
in the present work as its contribution is negligible in the threshold region we consider. 
For the neutron magnetic and the proton form factors we take the dipole fit while the
neutron electric form factor is taken from \cite{GaK71}.  

\begin{figure}[ht]

\centerline{\resizebox*{14cm}{8cm}{\includegraphics*[angle=0]{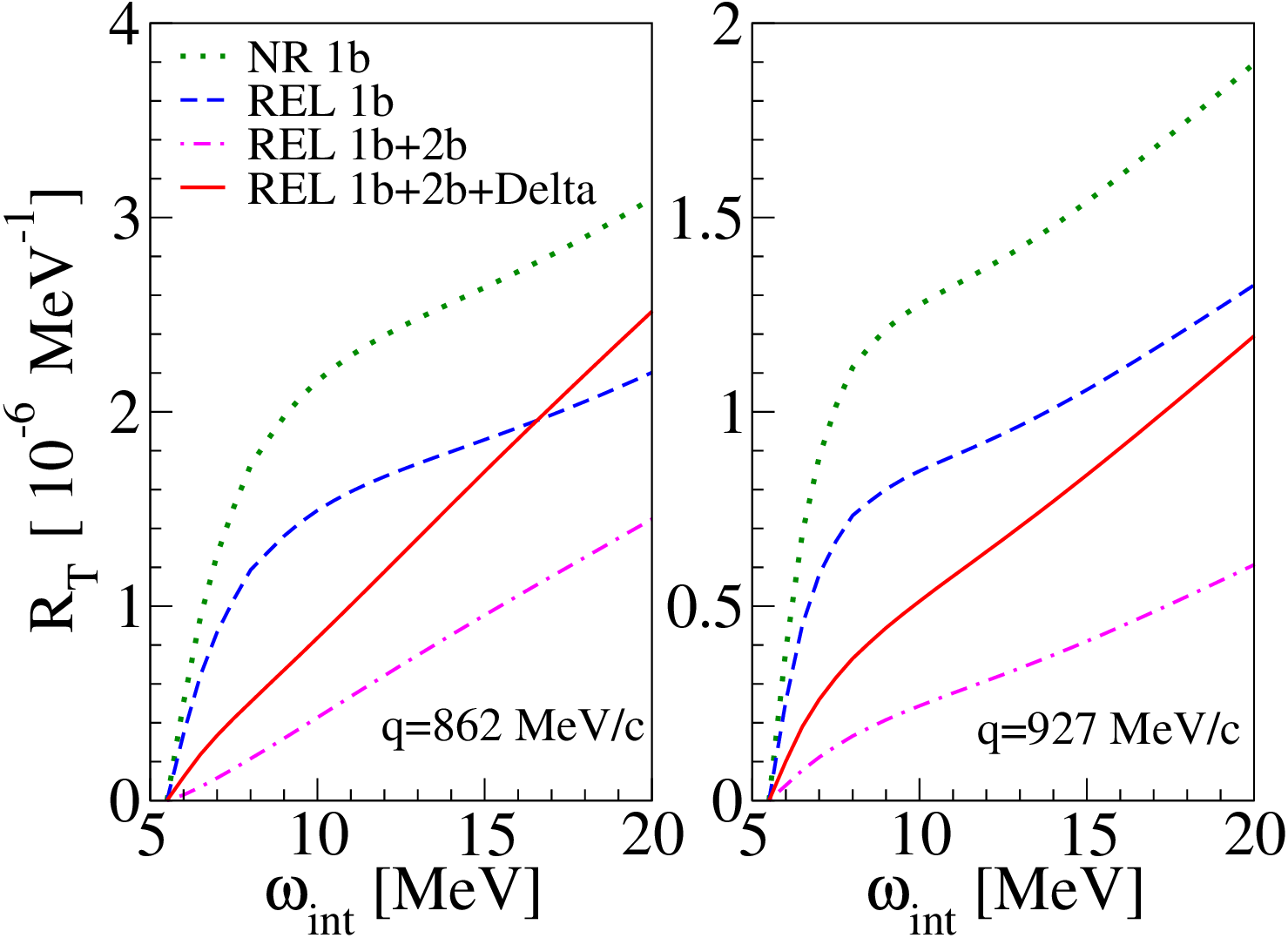}}}
\caption{Theoretical results for $R_T$ of $^3$He as function of internal excitation energy $\omega_{\rm int}$ at $q$=862
 and 927 MeV/c with various current operators: non-relativistic nucleon one-body (dotted),
with additional relativistic corrections (dashed), plus additional MEC (dash-dotted),
further addition of $\Delta$-IC (solid).}  \label{q862}
\end{figure}

Fig.~2 displays our $R_T$ results for several calculational options. The dominant
transition multipolarity contributing  at these near threshold energies is M1.
One sees that relativistic effects reduce the M1 transition strength considerably.
If in addition MEC are also taken into account then the M1 contribution drops markedly
leading to a rather different low-energy behavior of $R_T$. Inclusion
 of $\Delta$-IC restores some of this lost strength and demonstrates, as anticipated,
 that the $\Delta$ effect is quite large. 

Now we turn to a comparison of our results with those of \cite{Deltuva1}. For the comparison
one should keep in mind that there are differences between the two calculations.
Thus in \cite{Deltuva1}, (i) relativistic currents have not been considered, 
(ii) the Coulomb force is neglected in the final state interaction, 
(iii) their nucleonic potential model, the CDBonn \cite{CDBonn}, is different from ours
and does not reproduce the $^3$He binding energy
 and (iv) the full coupled channel calculation, CDBonn+$\Delta$ \cite{DeM03}, is
 a more consistent treatment of $\Delta$ degrees of freedom than the IA, but leads
 to a slight underestimation of the $^3$He binding energy. There is another point
 which makes the comparison a bit more difficult. Namely, the $R_T$ of \cite{Deltuva1} 
is not calculated for a constant momentum transfer, in fact $q$ is slightly decreasing
 with growing energy. Therefore, in Fig.~3, we prefer to display the results
 for each $q$
 in two panels. One sees that despite the various differences mentioned above
the $\Delta$-effects in both calculations are very similar. However, one also notes
that relativistic corrections lead to an opposite effect,
which is of the same size at $q=862$ MeV/c, but somewhat weaker at $q=927$ MeV/c. 
By comparison of results for $R_T$ at about $q=500$ MeV/c from \cite{LEOT10}  against those 
of \cite{Deltuva1} 
one finds again an at least partial cancellation of relativistic and $\Delta$ contributions close to the breakup threshold. 
The stronger increase of $R_T$ in the calculation of \cite{Deltuva1} at higher energies, 
seen in Fig.~3, partly originates 
from the non-constant  momentum transfers used in \cite{Deltuva1} as mentioned above
(see also discussion of Fig.~4). 

\begin{figure}[ht]
\centerline{\resizebox*{12.cm}{8.cm}{\includegraphics*[angle=0]{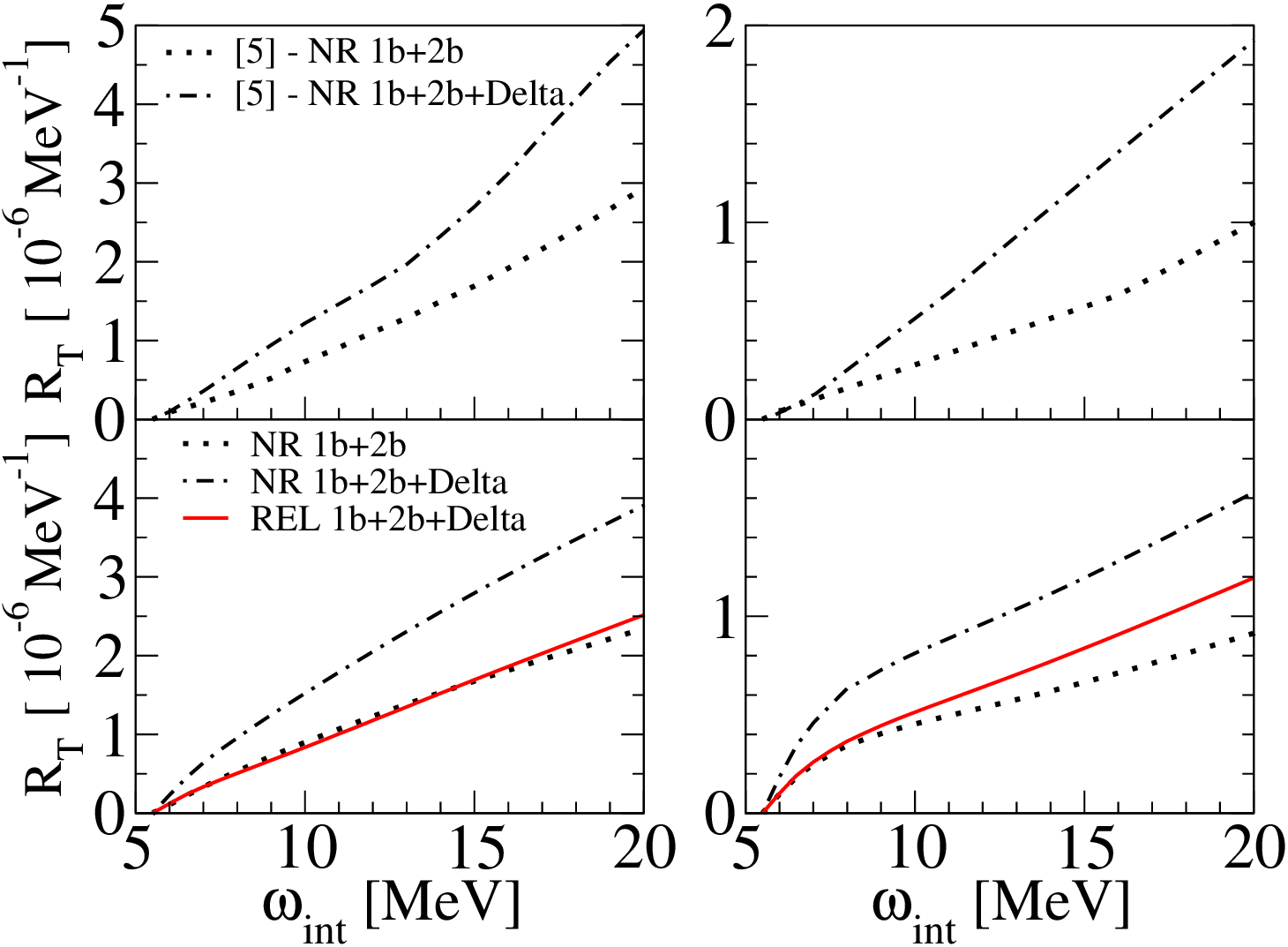}}}
\caption{$R_T$ of $^3$He.  Lower panels ($q=862$ MeV/c left, $q=927$ MeV/c right): theoretical result from present work with non-relativistic nucleon one-body current and MEC (dotted) and additional $\Delta$-IC
dash-dotted, further addition of relativistic corrections for nucleon one-body current
(solid). Upper panels ($q$ at threshold as in lower panels, but slightly varying with $\omega_{int}$, see text) : theoretical results
from Deltuva {\it et al.} \cite{Deltuva1} with non-relativistic nucleon one-body current and MEC (dotted)
and additional $\Delta$-contributions (dash-dotted).}  \label{q862fig2}
\end{figure}

Finally in Fig.~4 we compare the results of our calculation with experimental data \cite{Hi03}.
Again, as in the calculation of \cite{Deltuva1} the momentum transfer is only quasi-constant.
%Therefore we restrict the comparison to very low energies, where $q=862$ MeV/c is sufficiently
%correct. 
Therefore, in the left panel we show $R_T$ including all
current contributions for the two extreme $q$ values, i.e.
$q=862$ and $850$ MeV/c, and in addition for $q=862$ MeV/c the result where
only $\Delta$-IC are left out. The
lower $q$ corresponds to the data at about $\omega_{int}=20$ MeV, while the higher $q$ corresponds
to the threshold energy. In the right panel of the figure we
only show the results for $q=927$ MeV/c that corresponds to
the data close to threshold. The $\Delta$-IC contribution is seen to be essential for obtaining a good
agreement between theory and experiment 
below 10 MeV. However, at higher energies the increase due to $\Delta$-IC is not sufficient
to describe the data even if one considers the slight shift of $q$ with higher energies
represented by the dotted curve. The present case is rather similar to deuteron electrodisintegration
at higher momentum transfer, where at low excitation energy the leading M1 transition also
has a minimum. 
For the deuteron case it is known (see e.g. \cite{LeA83}) that various theoretical
ingredients, like for example the potential model dependence, can lead to rather large variations
of the theoretical result. Thus our present study cannot give a final answer concerning
the comparison of theory and experiment.

\begin{figure}[ht]
\centerline{\resizebox*{12.cm}{8.cm}{\includegraphics*[angle=0]{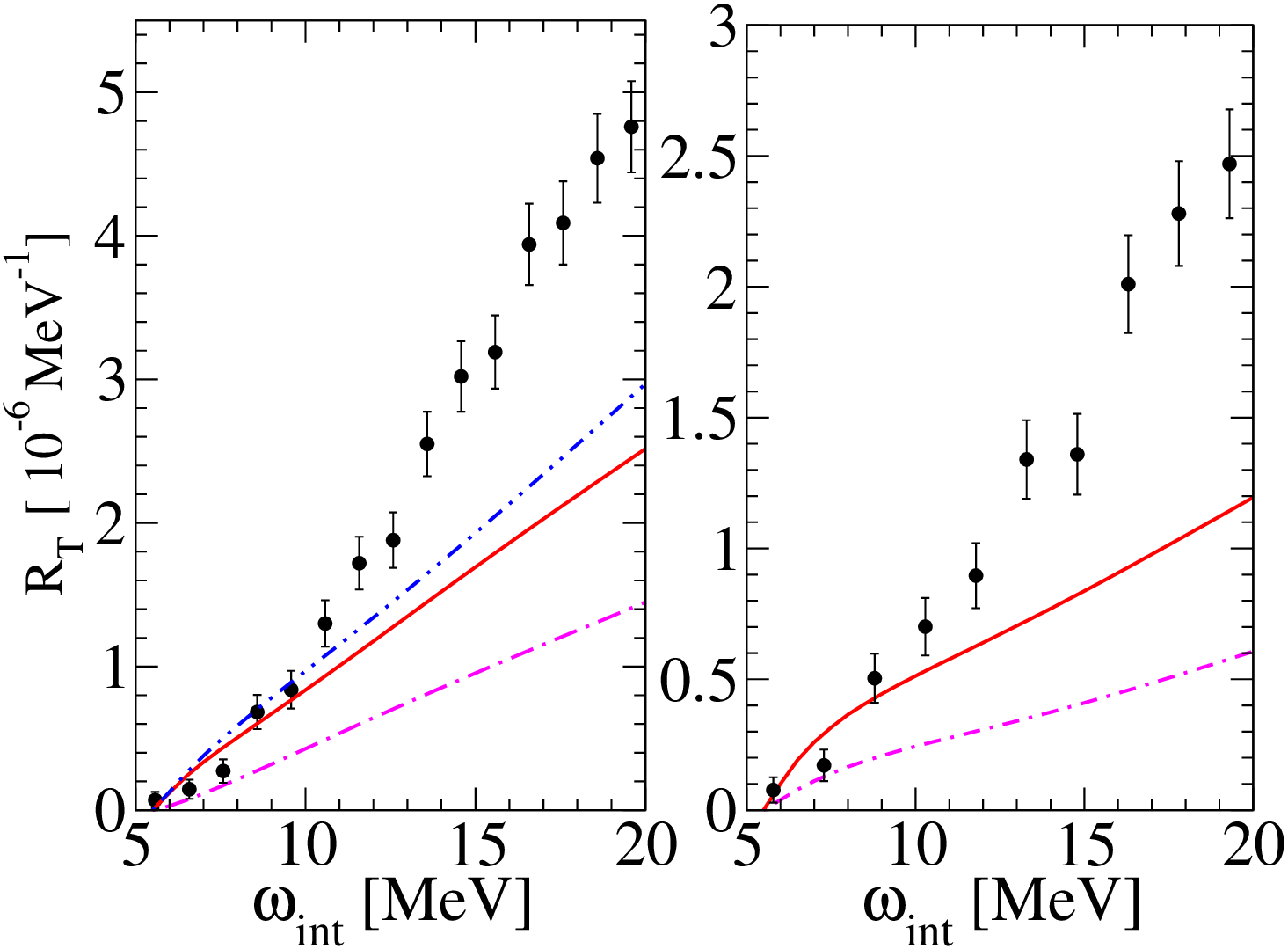}}}
\caption{$R_T$ of $^3$He for same kinematics and the same dash-dotted and solid curves
as in Fig.~2; 
in addition: result  with all current contributions at $q=$ 850 MeV/c
(dash-double dotted curve). 
Experimental data with slightly varying $q$-values from \cite{Hi03}.}  \label{q862fig3}
\end{figure}

We summarize our work as follows. We have illustrated how $\Delta$ degrees of freedom
are integrated into the LIT formalism for a calculation of the inelastic inclusive
transverse $(e,e')$ response function $R_T$ of $^3$He. The resulting coupled equations
for the Lorentz states of the $NNN$ and $NN\Delta$ channels contain,
as opposed to the corresponding coupled Schr\"odinger equation, source terms with
electromagnetic operators acting on the nuclear ground state. The $\Delta$ degrees
of freedom are present in three different forms: (i) in the potentials $V^{N'_1N'_2,N_1N_2}$,
(ii) in the $\Delta$-propagator, and (iii) in the current operators ${\bf j}_{N'_1N_1}$.
The coupled channel equation is solved in impulse approximation, where the $NNN$ and $NN\Delta$ channels are treated separately. First, the $NNN$ part is solved using a realistic nuclear
interaction with $NN$ and $NNN$ potentials. The result thus obtained is then used for the
solution of the $NN\Delta$ channel. The former gives a contribution to the electrodisintegration of a purely nucleonic final state, whereas the latter leads to a contribution to the pion
production channel. In the present work we have studied  $\Delta$-effects in $R_T$ of $^3$He
close to the breakup threshold at an momentum transfers of 
about 900 MeV/c. 
The response function
is affected by sizable MEC contributions, and, as in a previous full coupled channel
calculation \cite{Deltuva1} we find a considerable increase of $R_T$ due to $\Delta$ degrees of freedom. Unlike the calculation of \cite{Deltuva1} we here take into account
relativistic corrections to the nonrelativistic one-body current operator. 
At the kinematics considered here these relativistic corrections nearly cancel 
the $\Delta$-IC contribution. This cancellation in fact leads to
good agreement of our theoretical $R_T$ with experimental data 
at very low energy transfer, while the experimental $R_T$ is underestimated at
somewhat higher energies.

\section{Acknowledgment}

Acknowledgments of financial support are given to AuroraScience (L.P.Y.), to
the RFBR, grant 10-02-00718 and RMES, grant NS-7235.2010.2 (V.D.E.),
and to the National Science and Engineering Research Council of Canada (E.L.T.).

%\vfill\eject
%\newpage
\appendix
\section{Details of $g^c,h^c,g^d,h^d$} \label{gh}
The spin-isospin factors are
\begin{equation}
g ^c _{B'S'T',BS\mathcal{S}T\mathcal{T}} = \left \{ \begin{array}{c@{\quad \quad}l} \frac{1}{\sqrt{2}}f_{B'S'T',BS\mathcal{S}T\mathcal{T}} & \mbox{if $B'=1$ and $B = 0$, } \\ -(-)^{S+T}f_{B'S'T',BS\mathcal{S}T\mathcal{T}}\Big| _{s_1 \leftrightarrow s_2,t_1 \leftrightarrow t_2} & \mbox{if $B'=0$ and $B = 0$,} \\ 0 & \mbox{otherwise.} \end{array} \right. 
\end{equation}
\begin{equation}
g ^d _{B'S'T',BS\mathcal{S}T\mathcal{T}} = \left \{ \begin{array}{c@{\quad \quad}l} -(-1)^{S'+T'}\frac{1}{\sqrt{2}}f_{B'S'T',BS\mathcal{S}T\mathcal{T}} \Big| _{s_1' \leftrightarrow s_2',t_1' \leftrightarrow t_2'} & \mbox{if $B'=1$ and $B = 0$, } \\ -(-)^{S'+T'}\sqrt{2}f_{B'S'T',BS\mathcal{S}T\mathcal{T}}\Big| _{s_1' \leftrightarrow s_2',t_1' \leftrightarrow t_2'} & \mbox{if $B'=0$ and $B = 1$,} \\ 0 & \mbox{otherwise.} \end{array} \right. 
\end{equation}
\begin{equation}
f _{B'S'T',BS\mathcal{S}T\mathcal{T}} = (-)^{S+T}\hat{S'}\hat{S}\hat{T'}\hat{T}\left \{ {s_1' \; s_2'\; S' \atop s_3'\; \mathcal{S}\; S} \right \}\left \{ {t_1' \; t_2'\; T' \atop t_3'\; \mathcal{T}\; T} \right \} \delta _{s_1' s_3} \delta _{s_2' s_1} \delta _{s_3' s_2}.
\end{equation}
The spatial functions in coordinate representation are 
\begin{equation}
\langle {\bs \xi}_1{\bs \xi}_2 | h^c_{NK(Ll)\mathcal{LM_L}}\rangle = (-)^{L}\langle {\bs \xi}_1'({\bs \xi}_1, {\bs \xi}_2){\bs \xi}_2'({\bs \xi}_1, {\bs \xi}_2) | h_{NK(Ll)\mathcal{LM_L}}\rangle,
\end{equation}
\begin{equation}
\langle {\bs \xi}_1{\bs \xi}_2 | h^d_{NK(Ll)\mathcal{LM_L}}\rangle = \langle {\bs \xi}_1'(-{\bs \xi}_1, {\bs \xi}_2){\bs \xi}_2'(-{\bs \xi}_1, {\bs \xi}_2) | h_{NK(Ll)\mathcal{LM_L}}\rangle,
\end{equation}
where  ${\bs \xi}_1'({\bs \xi}_1, {\bs \xi}_2)$ and ${\bs \xi}_2'({\bs \xi}_1, {\bs \xi}_2) $ are connected to ${\bs \xi}_1$ and ${\bs \xi}_2 $ through the cycle operator $(123)$ by $\left \langle{\bf r}_1'{\bf r}_2'{\bf r}_3' s_1's_2's_3't_1't_2't_3'\right | = \left \langle{\bf r}_2{\bf r}_3{\bf r}_1s_2s_3s_1 t_2t_3t_1 \right | $, and using (\ref{xidef}) we have

\begin{eqnarray}
{\bs{\xi}}_1' ({\bs \xi}_1, {\bs \xi}_2)&=& -\sqrt{\frac{A_3A_1}{(A_2+A_3)(A_1+A_2)}}\;{\bs{\xi}}_1 + \sqrt{\frac{A_2(A_1+A_2+A_3)}{(A_2+A_3)(A_1+A_2)}}\;{\bs{\xi}}_2, \nonumber \\
{\bs{\xi}}_2' ({\bs \xi}_1, {\bs \xi}_2)&=& -\sqrt{\frac{(A_1+A_2+A_3)A_2}{(A_2+A_3)(A_1+A_2)}}\;{\bs{\xi}}_1 - \sqrt{\frac{A_1A_3}{(A_2+A_3)(A_1+A_2)}}\;{\bs{\xi}}_2 .
\end{eqnarray}

\medskip

\section {Kinetic Energy Calculational Details} \label{KE}

For a basis of antisymmetric states the kinetic energy can be written as
\begin{equation}
T = \frac{3}{2m_N}\frac{A_1+A_2}{A_1+A_2+A_3}{\bs \pi}^2_1,
\end{equation}
where ${\bs \pi}_1=-i\partial/\partial {\bs \xi}_1$ is the Jacobi
momentum conjugate to ${\bs \xi}_1$. 
Noting that for calculating the matrix elements of ${\bs \pi}^2_1$ between the basis
states (\ref{phik}) one may drop $\left(1-(12)\right)/\left[2(1+B''^2)\right]^{1/2}$,  we get
\begin{eqnarray}
\langle \varphi _{k'}|T| \varphi _k \rangle & = &\frac{3}{2m_N}
\delta _{\mathcal{L',L}}\delta _{\mathcal{S',S}}\delta _{\mathcal{J',J}}
\delta _{\mathcal{M_{J'},M_J}}\delta _{\mathcal{T',T}}
\delta _{\mathcal{M_{T'},M_T}}\sum_{B"S"T"}\frac{A_1^"+A_2^"}{A_1^"+A_2^"+A_3^"} \nonumber \\
 && \times \langle  F_{B"S"T",B'S'\mathcal{S}T'\mathcal{T}N'K'(L'l')
\mathcal{LM_L}}|{\bs \pi}^2_1|F_{B"S"T",BS\mathcal{S}T\mathcal{T}NK(Ll)
\mathcal{LM_L}} \rangle.
\end{eqnarray}
For the calculation of the spatial matrix elements we use the technique as
described in \cite{Ef02} to get
\begin{eqnarray}
\langle F_{j'\mathcal{L'M_{L'}} } | {\bs \pi}^2_1 | F_{j\mathcal{LM_L}} \rangle = \frac{\delta _{\mathcal {L',L}}{\delta _{\mathcal {M_{L'},M_L}}}8\pi ^2}{2\mathcal{L}+1}
 \int d \tau  _{int} \left \{   \sum _{\mathcal{M"}}  \left [ \frac{\partial}{\partial \xi _1} \underline{F_{j'\mathcal{LM"}}} \right ]
\left [ \frac{\partial}{\partial \xi _1} \underline{F_{j\mathcal{LM"}}} \right ]    \right .  \nonumber \\
\left .  + {\xi}^{-2}_1 \sum _{{\mathcal{M"}},\mu = \pm} \underline{l_{\mu}F_{j'\mathcal{LM"}}}\,\underline{l_{\mu}F_{j\mathcal{LM"}}}  \right \},
 \end{eqnarray}
where $d \tau _{int} = {\xi}^2_1 {\xi}^2_2 d{\xi}_1d{\xi}_2dt,\;t=\hat{{\bs \xi} _1}\! \cdot \! \hat{{\bs \xi} _2}$. The underlines mean that the space points take the value
\begin{equation}
\xi _{1x} = 0,\; \xi _{1y} = 0,\; \xi _{1z} = \xi _1,\; \xi _{2x} =\xi _2 \sqrt{1-t^2},\; \xi _{2y} = 0,\; \xi _{2z} = \xi _2t,\; \label{underlinepoint}
\end{equation}
the orbital angular momentum is given by ${\bf l} = {\bs \xi}_1 \times {\bs \pi}_1  $, and
\begin{equation}
\underline{l_{\pm 1}F_{j\mathcal{LM}}}  = -\frac{\xi _1}{\sqrt{2}}\left [(\partial / \partial \xi _{1x}) \pm i(\partial / \partial \xi _{1y}) \right ]F_{j\mathcal{LM}}
\end{equation}
with the derivatives taken at the space point of (\ref{underlinepoint}).

\medskip
\section{Calculational Details for $V(NN\to N\Delta)$ Potential} \label{pot}

In calculating matrix elements of the transition potentials between antisymmetric 
basis states one may omit the factor $(1-(12))/\sqrt(2(1+B'^2)$ 
from Eq.~(\ref{phik}) by using the substitutions
\[    \frac{1-(12)}{\sqrt{2(1+B^{'2})}}V^{N\Delta ,NN} \rightarrow   3 \cdot \sqrt{2} V_{12} (NN \rightarrow N\Delta), \] 
\[V^{NN,N\Delta}     \frac{1-(12)}{\sqrt{2(1+B^{2})}}\rightarrow 3 \cdot \sqrt{2}V_{12} (N\Delta \rightarrow NN). \]
Each basis state in (\ref{phik}) is the sum of several terms of the form
\[  \frac{1-(12)}{\sqrt{2(1+B^{'2})}}\left|\left(F_{j \mathcal{L}}\otimes \left(\left(s_1's_2'\right)S's_3'\right)\mathcal{S}\right)_{\mathcal{JM_J}}, \left(\left(t_1't_2'\right)T't_3'\right)\mathcal{TM_T}\right \rangle, \]
and this is also true for the basis of the $NNN$ part, but with $B = B' = 1,\;
(s_1,s_2,s_3) = (t_1,t_2,t_3)=(1/2,1/2,1/2)$, and 
$ \left(1-(12)\right)/\left[2(1+B'^2)\right]^{1/2} = 1$.
Therefore for the matrix elements of the operator $V_{12}$ we need 
\begin{eqnarray}
 \left \langle \left (F_{j'\mathcal{L'}}\otimes \left(\left(s_1's_2'\right)S's_3'\right)\mathcal{S'}\right)_{\mathcal{J'M_{J'}}} \right | V _0 (m_B) {\bs \sigma}_1 \! \cdot \!  {\bs \sigma}_2  \left | \left (F_{j\mathcal{L }}\otimes \left(\left(s_1 s_2 \right)S s_3 \right)\mathcal{S }\right)_{\mathcal{J M_{J }}} \right \rangle \nonumber \\ 
 = -\delta _{\mathcal{L',L}}\delta _{S',S}\delta _{s_3',s_3}\delta _{\mathcal{S',S}}\delta _{\mathcal{J',J}}\delta _{\mathcal{M_{J'},M_J}} \frac{1}{\hat{
{\mathcal{L}}}}\langle {F_{j'\mathcal{L}}}||V _0 (m_B)||{F_{j\mathcal{L}}}\rangle \nonumber \\
 \times (-1)^{1+S'+s_1+s_2'} \left \{ {s_1'\;s_1\;1 \atop s_2\;s_2'\;S'} \right \} \langle s_1' || \sigma _1 || s_1 \rangle \langle s_2' || \sigma _2 || s_2 \rangle.
\end{eqnarray}
In the spatial matrix elements entering here we note that the $NNN$ component
and the $NN\Delta$ component of the wave function are given in terms
of the Jacobi vectors of the same form  (\ref{xidef}) but with different mass numbers. 
To perform the integration one needs to express one set of the Jacobi vectors
in terms of the other via
\begin{eqnarray}
\overline{{\bs{\xi}}}_1 ({\bs \xi}_1, {\bs \xi}_2)&=& \sqrt{\frac{A_2'A_1'(A_1+A_2)}{(A_1'+A_2')A_1A_2}}\;{\bs{\xi}}_1 \nonumber \\
\overline{{\bs{\xi}}}_2 ({\bs \xi}_1, {\bs \xi}_2)&=& \frac{(A_1'A_2-A_2'A_1)}{\sqrt{A'(A_1'+A_2')A_1A_2(A_1+A_2)}}\;{\bs{\xi}}_1 + \sqrt{\frac{(A_1'+A_2')A_3'A}{A'(A_1+A_2)A_3}}\;{\bs{\xi}}_2, 
\end{eqnarray}
where $A' = A_1'+A_2'+A_3',\; A=A_1+A_2+A_3.$ The integration is done as 
\begin{equation}
\langle {F_{j'\mathcal{L}}}||V _0 (m_B)||{F_{j\mathcal{L}}}\rangle = 
8\pi ^2 (2\mathcal{L}+1)^{-1/2}\int d \tau _{int} \sum _ {\mathcal M}
\underline{F_{j'\mathcal{LM}}}(\overline{{\bs{\xi}}}_1,\overline{{\bs{\xi}}}_2)V _0 (m_B)\underline{F_{j\mathcal{LM}}}({\bs{\xi}}_1,{\bs{\xi}}_2).
\end{equation}

In addition we need ( using ${\mathcal N}^{(2)}_M=4(\pi/5)^{1/2}Y_{2M}({\bf n})$)
\begin{eqnarray}
 && \left \langle \left (F_{j'\mathcal{L'}}\otimes \left(\left(s_1's_2'\right)S's_3'\right)\mathcal{S'}\right)_{\mathcal{J'M_{J'}}} \right | V _2 (m_B)S_{12}   \left | \left (F_{j\mathcal{L }}\otimes \left(\left(s_1 s_2 \right)S s_3 \right)\mathcal{S }\right)_{\mathcal{J M_{J }}} \right \rangle \nonumber \\ 
 && \; \; \; \;   = \delta _{\mathcal{J',J}}\delta _{\mathcal{M_{J'},M_J}} (-1)^{{\mathcal {L+S'+J'}}}{3}/{2}\left \{ {\mathcal{L'}\;\mathcal{L}\; 2 \atop \mathcal{S}\;\mathcal{S'}\;\mathcal{J'} }\right \}
\langle {F_{j'\mathcal{L'}}}||V _2 (m_B){\mathcal N}^{(2)}||{F_{j\mathcal{L}}}\rangle \nonumber \\
&& \; \; \;  \qquad \times \langle \left(\left(s_1's_2'\right)S's_3'\right)\mathcal{S'} || \Sigma^{(2)} || \left(\left(s_1 s_2 \right)S s_3 \right)\mathcal{S } \rangle,
\end{eqnarray}

\begin{equation}
\langle {F_{j'\mathcal{L'}}}||V _2 (m_B){\mathcal N}^{(2)}||{F_{j\mathcal{L}}}\rangle \nonumber \\
 = 16\pi ^2 (2\mathcal{L'}+1)^{-1/2}\int d \tau _{int} \sum _ {\mathcal M}C^{\mathcal{L'M}}_{\mathcal{LM}20}
\underline{F_{j'\mathcal{L'M}}}(\overline{{\bs{\xi}}}_1,\overline{{\bs{\xi}}}_2)
V _2 (m_B)
\underline{F_{j\mathcal{LM}}}({\bs{\xi}}_1,{\bs{\xi}}_2),
\end{equation}
and
\begin{eqnarray}
 &&  \langle \left(\left(s_1's_2'\right)S's_3'\right)\mathcal{S'} || \Sigma^{(2)} || \left(\left(s_1 s_2 \right)S s_3 \right)\mathcal{S } \rangle \nonumber \\
&&  = \delta _{s_3',s_3} (-1)^{S'+s_3'+{\mathcal {S}}}\sqrt{30}/{3}\hat{S'}\hat{S}\hat{\mathcal{S'}}\hat{\mathcal{S}}\left \{ {\mathcal{S'}\;\mathcal{S}\; 2 \atop {S}\;{S'}\;{s_3'} }\right \} \left \{ \begin{array}{ccc}
s_1'\; & s_1\; & 1 \\  s_2' \; & s_2 \; &1 \\ S' \; & S \; & 2 \end{array} \right \} 
 \langle s_1' || \sigma _1 || s_1 \rangle \langle s_2' || \sigma _2 || s_2 \rangle.
\end{eqnarray}

\medskip
%\newpage
\section{$T_{jm}^l$ Multipoles of One-Body Currents Relating $\Delta$} \label{multidelta}
For the magnetic multipoles one has
\begin{eqnarray}
{T}^{j}_{jm}=\sum_i [{T}^{j,\,{\rm spin}}_{jm}(i) + 
{T}^{j,\,{\rm conv}}_{jm}(i)].
\end{eqnarray}
we have 
\begin{eqnarray}
\!\!\!\!{T}^{j,\,{\rm spin}}_{\Delta N,jm}(i)=\frac{1}{m_{\Delta}}\frac{q}{2}
\overline{\mu}^{\Delta N}_p\tau_{zi}
\left\{
\sqrt{\frac{j}{2j+1}}j_{j+1}(qr'_i)[Y_{j+1}(\hat{\mathbf{r}}'_i)\otimes
\sigma_i]_{jm}\right.  \nonumber \\ \!\!\!\!\!\!\!\left.-
\sqrt{\frac{j+1}{2j+1}}j_{j-1}(qr'_i)[Y_{j-1}(\hat{\mathbf{r}}'_i)\otimes
\sigma_i]_{jm}\right\},
\end{eqnarray}
\begin{eqnarray}
\!\!\!\!{T}^{j,\,{\rm spin}}_{\Delta \Delta,jm}(i)=\frac{1}{m_{\Delta}}\frac{q}{2}{\bar\mu}^{\Delta}_p\left( \frac{1}{2}+\frac{1}{2}\tau_{zi} \right)
\left\{
\sqrt{\frac{j}{2j+1}}j_{j+1}(qr'_i)[Y_{j+1}(\hat{\mathbf{r}}'_i)\otimes
\sigma_i]_{jm}\right.  \nonumber \\ \!\!\!\!\!\!\!\left.-
\sqrt{\frac{j+1}{2j+1}}j_{j-1}(qr'_i)[Y_{j-1}(\hat{\mathbf{r}}'_i)\otimes
\sigma_i]_{jm}\right\},
\end{eqnarray}
%{
\begin{eqnarray}
{T}^{j,{\rm conv}}_{\Delta \Delta,jm}(i)=\frac{1}{m_{\Delta}}
\left(\frac{1}{2}+\frac{1}{2}\tau_{zi}\right)j_j(qr'_i)\left[Y_j(\hat{\mathbf{r}
}'_i)\otimes
\partial'_i\right]_{jm}.
\end{eqnarray}
The quantity $\partial'_\mu$ is defined by the relationship $-i\partial'_\mu=p'_\mu$, and 
\[\partial'^{(3)}_\mu=
\left[\frac{(A_1+A_2)A_3}{A}\right]^{1/2}\frac{\partial}{\partial\xi_{2,\mu}}.\]

For the electric multipoles 
\begin{eqnarray}
T^{l}_{jm}=\sum_i [T^{l,\,{\rm spin}}_{jm}(i) + T^{l,\,{\rm conv}}_{jm}(i)]
\end{eqnarray}
where $l=j\pm 1$.
One obtains
\begin{eqnarray}
T^{j\pm 1,\,{\rm spin}}_{ \Delta N,jm}(i)=-\frac{1}{m_{\Delta}}\frac{q}{2}
\overline{\mu}^{\Delta N}_p\tau_{zi}
\sqrt{\frac{j+(1\mp
1)/2} {2j+1}}j_j(qr'_i)\left[Y_j
(\hat{\mathbf{r}}_i')\otimes \sigma_i\right]_{jm}
\end{eqnarray}
\begin{eqnarray}
T^{j\pm 1,\,{\rm spin}}_{ \Delta \Delta,jm}(i)=-\frac{1}{m_{\Delta}}\frac{q}{2}{\bar\mu}^{\Delta}_p\left( \frac{1}{2}+\frac{1}{2}\tau_{zi} \right)
\sqrt{\frac{j+(1\mp
1)/2} {2j+1}}j_j(qr'_i)\left[Y_j
(\hat{\mathbf{r}}_i')\otimes \sigma_i\right]_{jm}
\end{eqnarray}
\begin{eqnarray}
\label{tjconv}
T^{j\pm 1,\,{\rm conv}}_{\Delta \Delta,jm}(i)=\pm\frac{1}{m_{\Delta}}
\left(\frac{1}{2}+\frac{1}{2}
\tau_{zi}\right)\left\{j_{j\pm 1}(qr'_i)\left[Y_{j\pm 1}(\hat{\mathbf{r}}'_i)\otimes
\partial'_i\right]_{jm} \right.\nonumber
\\\left.-\kappa \frac{q}{2}\sqrt{\frac{j+(1\pm
1)/2}{2j+1}}j_{j}(qr'_i)Y_{jm}(\hat{\mathbf{r}}'_i)\right\}.
\end{eqnarray}

 For the combined $r$- and spin space operator we only need the reduced
 matrix element for the calculation of the  response function, as seen
 in (\ref{basisq}). The reduced matrix element  is given by
\begin{eqnarray}
 \left \langle \left (F_{j'\mathcal{L'}}\otimes \left(\left(s_1's_2'\right)
S's_3'\right)\mathcal{S'}\right)_{\mathcal{J'}} \right \|  (O_l \otimes O_s)_{j}
 \left \| \left (F_{j\mathcal{L }}\otimes \left(\left(s_1 s_2 \right)S s_3 
\right)\mathcal{S } \right)_{\mathcal{J }} \right \rangle \nonumber \\  
= \hat{\mathcal{J'}}\hat{\mathcal{J}}\hat{j} \left \{ \begin{array}{ccc}
{\mathcal{L'}}\; & {\mathcal L} \; & l \\  {\mathcal{S'}}\; & {\mathcal S} 
\; & s  \\ {\mathcal{J'}}\; & {\mathcal J} \; & j  \end{array} \right \}
\langle {F_{j'\mathcal{L'}}}|| O_l ||{F_{j\mathcal{L}}}\rangle
   \langle \left(\left(s_1's_2'\right)S's_3'\right)\mathcal{S'} || O_s || \left(
\left(s_1 s_2 \right)S s_3 \right)\mathcal{S } \rangle,
\end{eqnarray}
where 
\begin{equation}
\langle {F_{j'\mathcal{L'}}}|| O_l ||{F_{j\mathcal{L}}}\rangle
 \nonumber \\ = 8\pi ^2 (2\mathcal{L'}+1)^{-1/2}\int d \tau _{int}
 \sum _ {{\mathcal M}m}C^{{\mathcal{L'(M}}+m)}_{\mathcal{LM}lm}
\underline{F_{j'\mathcal{L'M}}}(\overline{{\bs{\xi}}}_1,\overline{{\bs{\xi}}}_2)
\underline{O_l F_{j\mathcal{LM}}}({\bs{\xi}}_1,{\bs{\xi}}_2),
\end{equation}
and $O_l$ is a function of relative coordinate and momentum \[ {\bf r}_3-{\bf R}_{cm} =  \left[ \frac{(A_1+A_2)}{AA_3}\right]^{1/2}{\bs \xi}_{2},\qquad
i\left({\bf p}_3-\frac{A_3}{A}{\bf P}_{cm}\right)=\left[\frac{(A_1+A_2)A_3}{A}\right]^{1/2}\frac{\partial}{\partial{\bs \xi}_{2}}. \]

\end{document}